\documentclass{ws-ijmpd}
\usepackage[super,compress]{cite}
\begin{document}


%
\catchline{}{}{}{}{}
%

\title{The Puzzle of Muons in Extensive Air Showers
}

\author{Maciej Rybczy\'{n}ski}

\address{Institute of Physics, Jan Kochanowski University\\ 
Kielce 25-406, Poland\\
maciej.rybczynski@ujk.edu.pl}

\author{Zbigniew W\l odarczyk}

\address{Institute of Physics, Jan Kochanowski University\\ 
Kielce 25-406, Poland\\
zbigniew.wlodarczyk@ujk.edu.pl}

\maketitle

\begin{history}
\received{Day Month Year}
\revised{Day Month Year}
\end{history}

\begin{abstract}
In order to examine a muon excess observed by the Pierre Auger Observatory, detailed Monte Carlo simulations
were carried out for primary protons, iron nuclei and strangelets (hypothetical stable lumps of strange quark matter).
We obtained a rough agreement between the simulations and the data for ordinary nuclei without any
contribution of strangelets in primary flux of cosmic rays.
Our simulations suggest that the shower observables are dominated by details of hadronic interaction models.


\end{abstract}

\ccode{PACS numbers: 96.50.sd, 13.85.Tp, 98.70.Sa,  24.85.+p}


\section{Introduction}
\label{sect:intro}
Ultra-high energy ($E>10^{18}$~eV) cosmic rays (UHECR) provide a formidable beam to study particle collisions at center-of-mass energies and kinematical regimes not accessible at terrestrial accelerators~\cite{Anchordoqui:1998nq}. Understanding the mass composition of UHECR at Earth is fundamental to unveil their production and propagation mechanisms. The identity of the highest-energy cosmic rays is still an open question. A possible dominance of protons or iron nuclei in the cosmic ray flux poses problems~\cite{Schwarzschild:2010zzc}. Ultra-high energy cosmic rays can only be observed indirectly, through air showers. Seeking to determine the nuclear identities of the UHECR particles, the development of extensive air showers (EAS) of secondary particles in the atmosphere was extensively examined. The mass composition of cosmic rays can be derived from certain air shower observables, but the inference is limited by our theoretical understanding of the air shower development. Air shower simulations require knowledge of hadronic interaction properties at very high energies and in phase space regions that are not well covered by accelerator experiments. The systematic uncertainty of the inferred mass composition can be reduced by studying different observables. The slant depth of the shower maximum $X_{max}$ is a prominent mass-sensitive tracer, since it can be measured directly with fluorescence telescopes.
It is well known (cf.~\cite{AlvarezMuniz:2002ne,AlvarezMuniz:2004bx,Risse:2004ac,Ulrich:2009zq}) that most of the charged particles in the shower are electrons and positrons with energies near the critical energy ($\epsilon $=81 MeV in air) coming from electromagnetic sub-showers initiated by $\pi ^{0}$-decay photons. The average depth of maximum for an electromagnetic shower initiated by a photon with energy $E_{\gamma}$ is 
\begin{equation}
\langle X^{em}_{max}\left(E_{\gamma}\right)\rangle = X_{0} \ln(E_{\gamma}/\epsilon),
\end{equation}
where $X_{0} = 37~{\rm g/cm^2}$ is the radiation length in air.

A nuclear-initiated shower consists of a hadronic core feeding the electromagnetic component primarily through $\pi ^0$ production. In general, for an incident nucleus of mass A and total energy $E_{0}$, including protons with $A$=1, the average depth of shower maximum is expressed as
\begin{equation}
\langle X_{max}\left(E_{0}\right)\rangle=
\langle X^{em}_{max}\left[\left(E_{0}/A\right)\left(K/\langle N\rangle\right)\right]\rangle+\langle X_{1}\rangle,
\end{equation}
where $\langle X_{1}\rangle$ is the mean depth of the interaction with maximal energy deposition within the  shower (usually called the depth of the first interaction), $K$ denotes inelasticity and $\langle N\rangle$ is related to the multiplicity of secondaries in the high-energy hadronic interactions in the cascade. 
If the composition changes with energy, then $A$ depends on energy and $\langle X_{max}\rangle$ changes accordingly. The situation is, however, essentially more complicated. Whereas for a primary nucleus in which the energy is to a good approximation simply divided into $A$ equal parts, in a hadronic cascade, instead, there is a hierarchy of energies of secondary particles in each interaction, and a similar (approximately geometric) hierarchy of interaction energies in the cascade. In this case $\langle N\rangle$ has to be understood as some kind of {\it effective} multiplicity, which does not have a straightforward definition in general. For this reason the variation of the primary composition, or the violation of Feynman scaling have been widely discussed from many years. In addition, the inelasticity $K$ can also change with energy. 

The number of muons in an air shower is another powerful tracer of the mass. Simulations show that the number of produced muons, $N_{\mu}$, rises almost linearly with the cosmic-ray energy $E$, and increases with a small power of the cosmic-ray mass $A$. This behaviour can be understood in terms of the generalized Heitler model of hadronic air showers~\cite{Matthews:2005sd}, which predicts
\begin{equation}
N_{\mu}=A^{1-\beta}\left(E/\xi_{c}\right)^\beta,
\label{eq:heit}
\end{equation}
where $\xi_{c}$ is the critical energy at which charged pions decay into muons, and $\beta \approx$ 0.9. Detailed simulations show further dependences on hadronic interaction properties, such as multiplicity, inelasticity, charge ratio and baryon-antibaryon pair production~\cite{Pierog:2006qv,Ulrich:2010rg}. The dependence of the muon number $N_{\mu}$ on the mass of the primary cosmic rays is complementary to the depth of the shower maximum $X_{max}$. If both observables are combined, the internal consistency of the mass composition can be tested. Over the past few decades, it has been suspected that the number of registered muons at the surface of the Earth is  tens of percentage points higher than what it should be, according to existing hadronic interaction models~\cite{AbuZayyad:1999xa,Aab:2014pza}. Recently, a study from the Pierre Auger Collaboration (Auger) has strengthened this suspicion, using a novel technique to mitigate some of the measurement uncertainties of earlier methods~\cite{Aab:2014pza,Aab:2016hkv}. The new analysis of Auger data suggests that the hadronic component of showers with primary energy $E>10^{18}$~eV contains about 30\% to 60\% more muons than expected. Also the number of muons with energies above 0.75~GeV, determined by the Sydney University Giant Air-shower Recorder (SUGAR), exceeds the simulated one by the factors $\sim 1.67$ and $\sim 1.28$ for $10^{17}$~eV proton and iron primaries, respectively. The muon excess grows moderately with the primary energy, increasing by an additional factor of 1.2 for $10^{18.5}$ eV primaries~\cite{Bellido:2018toz}~\footnote{However, Moscow State University Extensive Air Shower (EAS-MSU) data suggest that no muon excess is seen in the inner part of EAS induced by $E \sim 10^{17}$~eV primaries~\cite{Fomin:2016kul}.}. 

To explain the muon excess, several new models have been proposed, exploring new physics~\cite{Farrar:2013sfa,Allen:2013hfa,AlvarezMuniz:2012dd} or new forms of matter, namely strange quark matter (SQM)~\cite{Anchordoqui:2016oxy}. Roughly, two possibilities arise: either we are dealing with novel particles or known particles have novel properties of multiparticle production processes.
 
In this paper, in addition to standard nuclei, a much heavier bulk of matter is considered as primaries inducing air showers. We adopt a purely phenomenological approach to develop an SQM scheme. In sharp contrast to previous models our approach is based on the assumption that ultra-high energy cosmic rays are very heavy strange quark objects, i.e. strangelets (see~\ref{app:stran} for more details). Using simple hadronic interaction model (see~\ref{app:model_desc} for details) which is suitable for modifications, the shower development was simulated using Monte Carlo technique and compared with the Pierre Auger Observatory results.

\section{EAS simulations}
\label{sect:eas_sim}

\begin{figure}[t]
\begin{center}
\includegraphics[angle=0,width=0.8\textwidth]{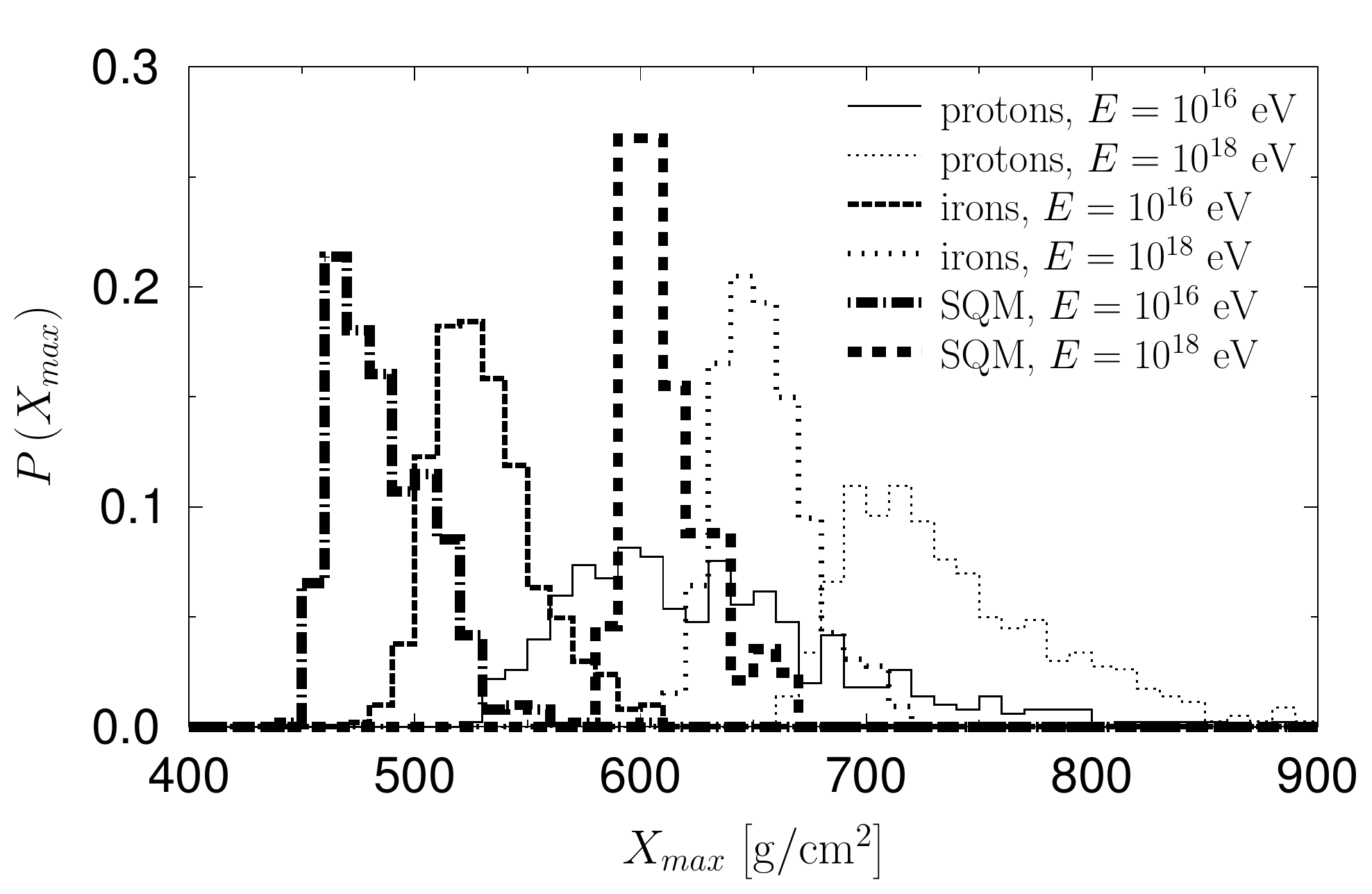}
\end{center}
\caption{\small Distributions of the position of maxima of showers generated by primary protons, iron nuclei, and SQM at primary energies $E=10^{16}$~eV and $E=10^{18}$~eV.}
\label{fig:max_top}
\end{figure}

\begin{figure}[t]
\begin{center}
\includegraphics[angle=0,width=0.8\textwidth]{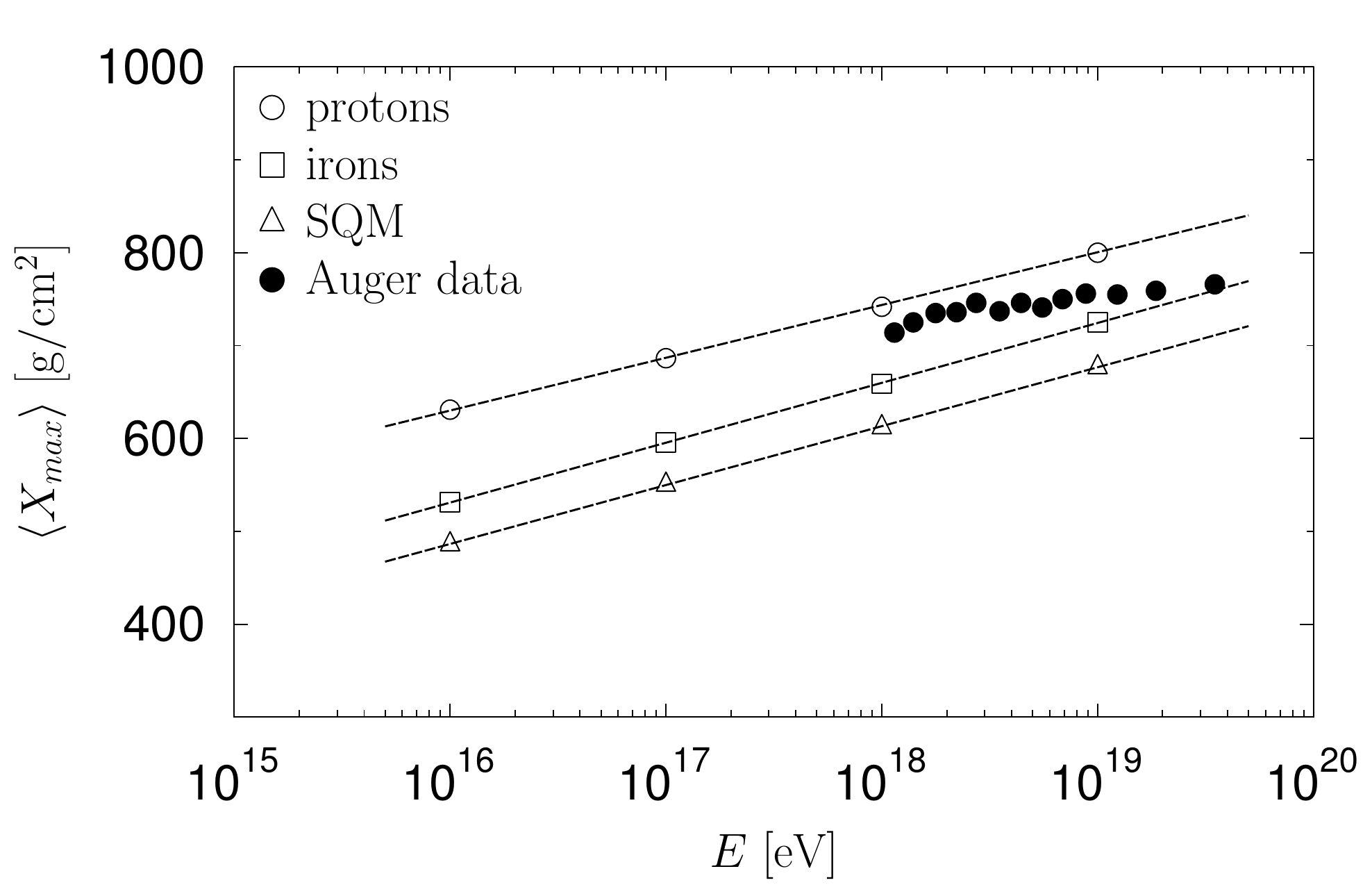}
\end{center}
\caption{\small Average position of the shower maximum as a function of the primary energy for EAS generated by protons, iron nuclei, and SQM. Dotted lines show a fit of the type $\langle X_{max}\rangle=c+d\cdot \ln \left(E/10^{19}~{\rm eV}\right)$. Pierre Auger Collaboration measurements~\cite{Abraham:2010yv} are indicated by black circles. See text for details.}
\label{fig:x_max_1}
\end{figure}

For the simulation of the propagation of extensive air showers in the Earth's atmosphere we have used a suitably modified SHOWERSIM~\cite{SHOWERSIM84} modular software. We performed Monte Carlo simulations of the EAS generated by primary nuclei (protons and  iron nuclei) and by primary strangelets with mass $A>320$ taken from the $A^{-7.5}$ distribution, for energies in the interval $10^{16}<E<10^{19}$~eV. In the analysis we have focused on the muons in the nuclear cascade with energies larger than $0.3$~GeV, which is the Cherenkov threshold for muons in water, reaching the Pierre Auger South Laboratory surface detector placed at the altitude $1425$~m above sea level. This corresponds to a total atmospheric depth of $X_{atm}= 750~{\rm g/cm^{2}}$ for the showers initiated at $\Theta=0~\deg$.

In figure~\ref{fig:max_top} we plot the distributions of the position of showers maxima generated by primary protons, iron nuclei and SQM. The average positions of showers maxima as a function of primary energy are shown in figure~\ref{fig:x_max_1} together with corresponding Pierre Auger Collaboration measurements~\cite{Abraham:2010yv}. We note a very good agreement between presented Auger results and our simulations. Obviously, the simulated values for different primaries follow a logarithmic trend, thus were fitted using the formula $\langle X_{max}\rangle=c+d\cdot \ln \left(E/10^{19}~{\rm eV}\right)$, with $c=801.4,~724.3,~683.8$ and $d=24.7,~28.0,~27.7$ for protons, iron nuclei and SQM, respectively.

\section{Results}
\label{sect:res}

\begin{figure}[t]
\begin{center}
\includegraphics[angle=0,width=0.8\textwidth]{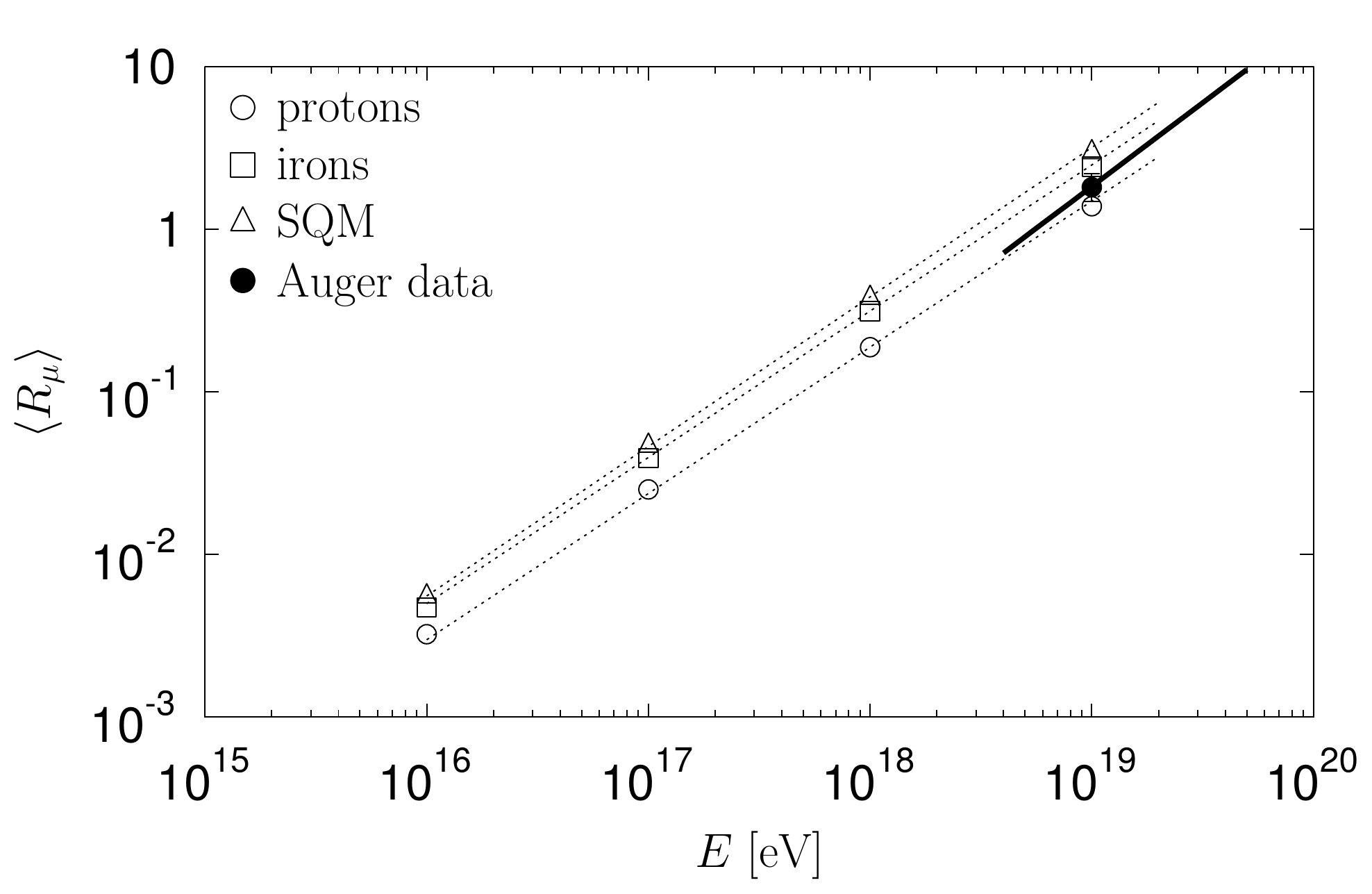}
\end{center}
\caption{\small Average muon content $\langle R_{\mu}\rangle$ of individual showers generated by primary protons, iron nuclei and SQM plotted as a function of primary energy $E$ together with the Pierre Auger Collaboration result taken from~\cite{Aab:2014pza}. The parametrizations given by equation~\ref{eq:mean_R} are plotted with the lines. The full thick line shows the Pierre Auger Collaboration parametrization while the thin dotted lines show our parametrization for protons, iron nuclei, and SQM. See text for details.}
\label{fig:n_mu_e}
\end{figure}

\begin{figure}[t]
\begin{center}
\includegraphics[angle=0,width=0.8\textwidth]{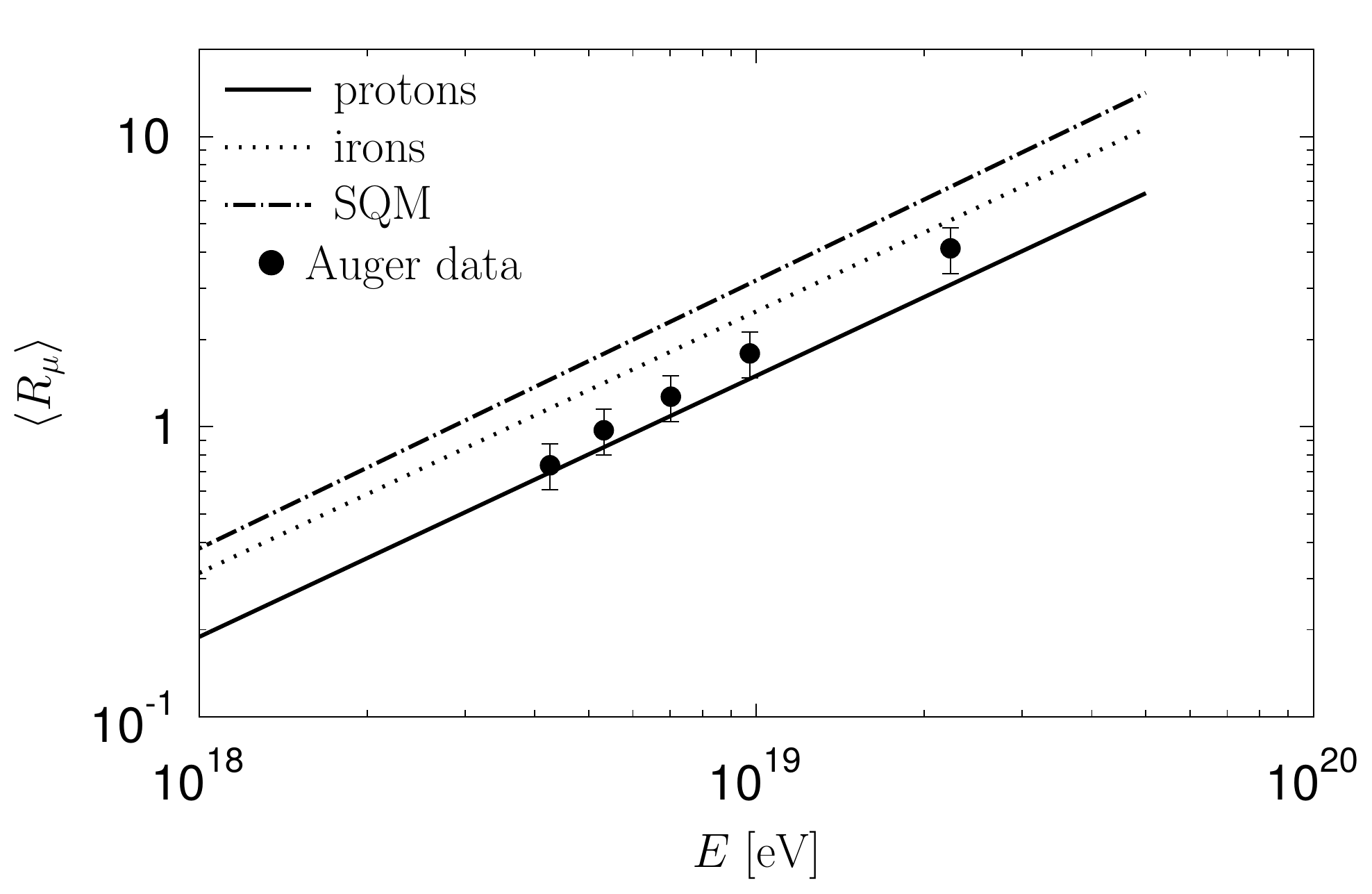}
\end{center}
\caption{\small The power-law fit to the SHOWERSIM simulations presented in figure~\ref{fig:n_mu_e} of the average muon contents $\langle R_{\mu}\rangle$ of individual showers generated by primary protons, iron nuclei, and SQM, plotted as a function of the primary energy $E$, together with the Pierre Auger Collaboration results taken from~\cite{Aab:2014pza}. See text for details.}
\label{fig:n_mu_1}
\end{figure}

In this section we provide a comparison of our SHOWERSIM simulations with the Pierre Auger Collaboration results presented in reference~\cite{Aab:2014pza}. We focus at the inclined showers generated with the average zenith angle $\langle \Theta\rangle=67~\deg$ (primaries with zenith angle $62 < \Theta <80~\deg$ were sampled). The Pierre Auger Collaboration presents the measured number of muons in inclined air showers using the scaled factor, relating the observed  muon densities at the ground to the average muon density of simulated EAS induced by protons at a fixed energy $10^{19}$~eV. The used scale factor is independent from the zenith angle and from the details of the location of the detector~\cite{Ave:2000dd,ThePierreAuger:2013eja}. Following~\cite{Aab:2014pza} the muon content $R_{\mu}$ is defined as:
\begin{equation}
R_{\mu}=\frac{N_{\mu}}{N_{\mu,19}},
\label{eq:R}
\end{equation}
where $N_{\mu}$ is the total number of muons at the ground in EAS generated by primary cosmic rays at different primary energies, and $N_{\mu,19}$ is the total number of muons in EAS generated by primaries with energy $E=10^{19}$~eV. Using equation~\ref{eq:heit} and following~\cite{Aab:2014pza} we used the power-law parametrization:
\begin{equation}
\langle R_{\mu}\rangle= a\cdot\left(E/10^{19}~{\rm eV}\right)^{b}
\label{eq:mean_R}
\end{equation}
with parameters $a=1.841$ and $b=1.029$ fitting Auger experimental events above $4\cdot 10^{18}$~eV. At zenith angle $\Theta=67~\deg$ the muon content $R_{\mu}=1$ corresponds to $N_{\mu}=1.455\cdot 10^{7}$ muons at the ground with energies above $0.3$~GeV. For model comparisons, as described in~\cite{Aab:2014pza}, the simulated number of muons should be then divided by $N_\mu=1.455\cdot 10^{7}$ to obtain $R_\mu$, which can be directly compared to Auger measurements. In figure~\ref{fig:n_mu_e} we show the simulated muon content $\langle R_{\mu}\rangle$ of individual showers generated by primary protons, iron nuclei, and SQM, as a function of the primary energy $E$. The Auger parametrization given by equation~\ref{eq:mean_R} is plotted with a thick line. 

In figure~\ref{fig:n_mu_1} we show the power-law fits (prepared using equation~\ref{eq:mean_R} with the parameters $a$ and $b$ equal to 1.5 and 0.9, 2.5 and 0.9, 3.2 and 0.92 for protons, irons, and SQM, respectively) to our simulated events using the SHOWERSIM generator for the average muon contents obtained in EAS initiated by protons, iron nuclei, and SQM, plotted as a function of the primary energy. Note that the parameters obtained in our SHOWERSIM simulations are very close to those obtained from Auger experiment, see reference~\cite{Aab:2014pza}. We also show in figure~\ref{fig:n_mu_1} the corresponding Auger measurement~\cite{Aab:2014pza}. It is noticeable that Auger results are located between the simulation of protons and iron nuclei. At lower energies, the experimental points are close to the $\langle R_{\mu}\rangle$ values obtained in the simulations assuming protons as primaries. However, at higher energies, the measured muon content approaches the $\langle R_{\mu}\rangle$ values coming from simulations of EAS generated by iron nuclei.


\begin{figure}[t]
\begin{center}
\includegraphics[angle=0,width=0.8\textwidth]{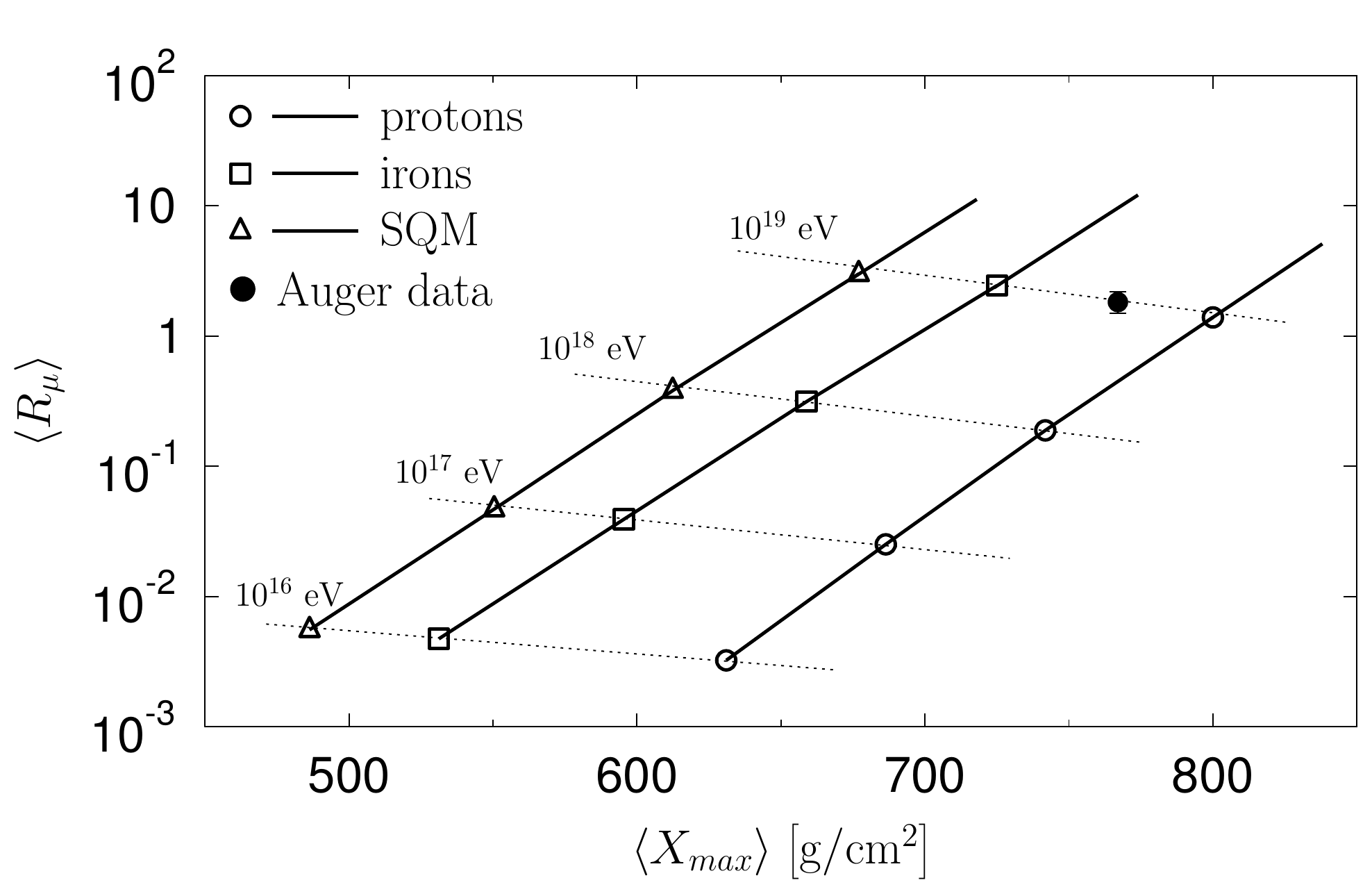}
\end{center}
\caption{\small Average muon content $\langle R_{\mu}\rangle$ of individual showers generated by primary protons, iron nuclei, and SQM as a function of the average position of the shower maximum. The full circle shows the Auger result~\cite{Aab:2014pza}. Dotted lines connect simulations performed at the same primary energy.}
\label{fig:x_max}
\end{figure}
 
The dependence of the average muon content $\langle R_{\mu}\rangle$ on the position of the showers maximum is presented in figure~\ref{fig:x_max}. It is remarkable that no heavier component than iron is needed to describe the experimental values of $\langle R_{\mu}\rangle$ at $10^{19}$~eV. We can explain experimental data without strange quark matter.

The muon content in the air shower, $R_{\mu}$, is a quantity related to the atomic mass $A$ of the primary cosmic ray. A possible implication for the mass composition is demonstrated on the example, taking into account the obtained experimentally $\langle R_{\mu}\rangle=1.82\pm 0.38$ and the relative standard deviation $\omega=\sigma\left(R_{\mu}\right)/\langle R_{\mu}\rangle=0.20\pm0.01$~\cite{Aab:2014pza}. For a single component, the values of $\omega$ can vary between $\omega=0.04$ for pure iron nuclei and $\omega=0.13$ for protons. More than two components (proton and iron) are needed to describe the first two moments of the $R_{\mu}$ distribution. The best description of the data is obtained with four components ($40\%$ protons, $20\%$ helium, $35\%$ nitrogen, and $5\%$ iron nuclei) while the addition of more species does not improve the quality of the fit. The comparatively small abundance of iron nuclei has been evaluated from the analysis of the $\langle X_{max}\rangle$~\textendash~$\sigma\left(X_{max}\right)$ difference~\cite{Wilk:2011ia,Wilk:2010iz}. Our results suggest roughly $\langle \ln A\rangle=1.4$ with standard deviation $\sigma\left(\ln A\right)=1.3$ at energies above $4\cdot 10^{18}$~eV.

\section{Concluding remarks}
\label{sect:cr}
\begin{figure}[t]
\begin{center}
\includegraphics[angle=0,width=0.8\textwidth]{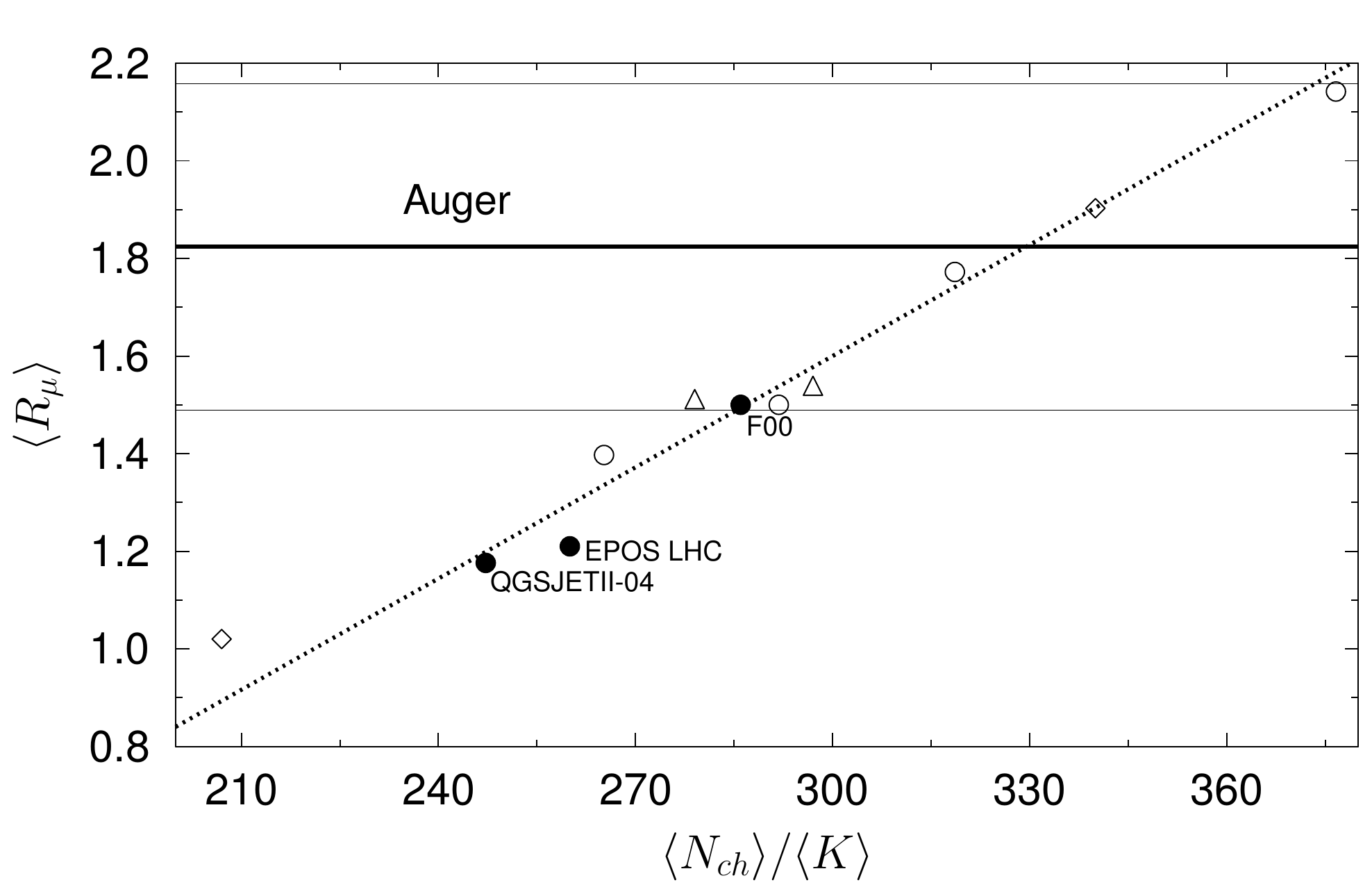}
\end{center}
\caption{\small Average muon content $\langle R_{\mu}\rangle$ of individual showers generated by primary protons at $10^{19}$~eV plotted as a function of charged multiplicity to inelasticity ratio $\langle N_{ch}\rangle/\langle K\rangle$ for our F00 model. EPOS LHC and QGSJETII-04 predictions are taken from~\cite{Pierog:2017nes}. The Pierre Auger Collaboration result is indicated by a thick full line, whereas the thin full lines show the uncertainty of the measurement, taken from~\cite{Aab:2014pza}. The dotted line shows a linear fit given by equation~\ref{eq:Rmu_fit}. Open symbols show the results obtained for different modifications of $\langle N_{ch}\rangle$ in the F00 model. See text for details.}
\label{fig:n_mu_mod}
\end{figure}

Very surprisingly the F00 model included in the SHOWERSIM simulation package~\cite{SHOWERSIM84} describe nicely the  muon content in EAS. Ordinary nuclei, without any contribution from strange quark matter in the primary flux of cosmic rays, can describe experimental data. Even if the strangelets contribute with a small amount in the primary flux and generate high multiplicity muon bundles, as we advocated recently~\cite{Kankiewicz:2016dha}, their influence on the average muon content $\langle R_{\mu}\rangle$ in EAS is negligible.

\begin{figure}[t]
\begin{center}
\includegraphics[angle=0,width=0.8\textwidth]{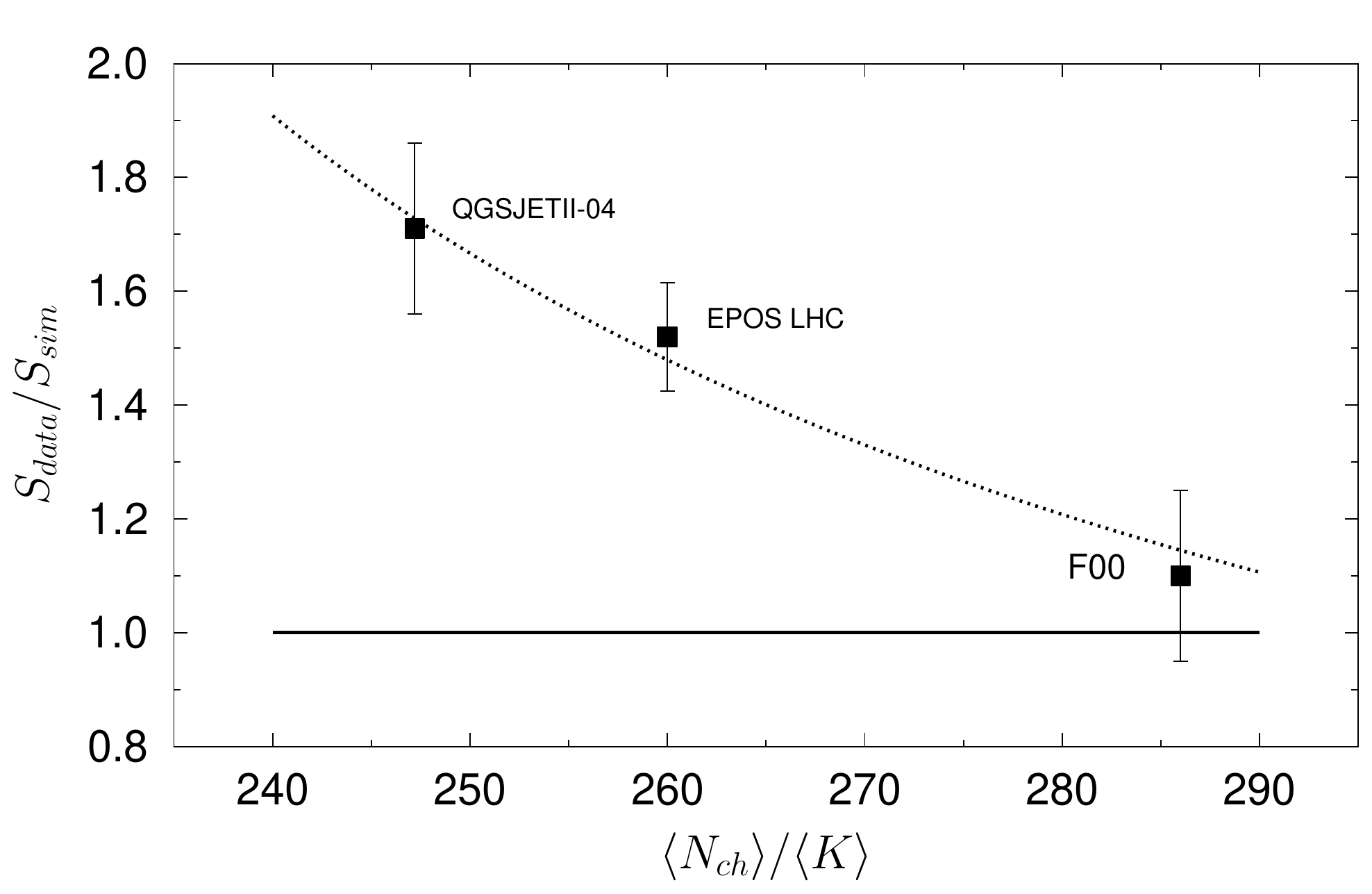}
\end{center}
\caption{\small Average ratio of $S\left(1000\right)$ at $\sec\left(\Theta\right)\simeq 1.95$ for observed and simulated events generated by primary protons at $10^{19}$~eV plotted as a function of $\langle N_{ch}\rangle/\langle K\rangle$. Model F00 prediction is compared with the EPOS LHC and QGSJETII-04 ones taken from~\cite{Aab:2016hkv}. Dotted line shows dependence given by $S_{data}/S_{sim}=\left(-1.3+0.0076 \langle N_{ch}\rangle/\langle K\rangle\right)^{-1}$.}
\label{fig:S_mu}
\end{figure}

Apparently, our model differs from modern high-energy hadronic models such as EPOS LHC or QGSJETII-04, which cannot fit the experimental values of $\langle R_{\mu}\rangle$. Comparing different characteristics, we find that in the discussed models, $\langle R_{\mu}\rangle $ and the ratio of charged multiplicity to inelasticity $\langle N_{ch}\rangle/\langle K \rangle$ follow the same ordering. This is illustrated in figure~\ref{fig:n_mu_mod}. In addition to standard F00 model (shown by full circle) we present results for different modified $\langle N_{ch}\rangle$ according to equation~\ref{eq:F00_ident_part} (shown by open symbols: circles correspond to different $\beta\left(K\right)$ and $\beta\left(\pi^{0}\right)$, triangles corresponds to different $\beta\left(K\right)$ and diamonds correspond to different $\beta\left(\pi^{0}\right)$). Roughly the number of muons depends linearly on the charged particle multiplicity:
\begin{equation}
\langle R_{\mu}\rangle=0.0076 \langle N_{ch}\rangle/\langle K\rangle - 0.68.
\label{eq:Rmu_fit}
\end{equation}

The lateral distribution of muons provides an additional valuable tool for testing hadronic interactions~\cite{Aab:2016hkv}. Figure~\ref{fig:S_mu} shows the ratio of $S\left(1000\right)$, the ground size at 1000 m from the shower core, for the Auger events~\cite{Aab:2016hkv} measured with its surface detector relative to that predicted for simulated events for F00 model. At large zenith angles, where the ground size, $S\left(r\right)$ is dominated by the muonic component, the simulated signal deficit roughly  corresponds to the muon content, $\langle R_{\mu}\rangle$ behaviour discussed above.

We have shown in our simple model that we are able to reproduce the muon content in EAS. The agreement between predicted and observed muon densities does not mean that F00 model describe properly multiparticle production processes. We do not claim that F00 is a final one. We use this simple model (suitable for easy modifications) only to demonstrate a method which can lead to understanding the observed muon excess. Models usually have many parameters and result in different muon densities. The question is what combination of parameters give the systematic changes of muon numbers. Comparing different sets of parameters we find that in the discussed models (EPOS LHC, QGSJETII-04, and F00) the muon content and the ratio of charged particle multiplicity to inelasticity follow the same ordering. Thus, the charged particle multiplicity to inelasticity ratio seems to be crucial for understanding the muon excess. Probably other combinations of parameters can also show a similar behavior. Following such a way we can find explanation of the observed muon excess. Clearly, further studies are required to understand the origin of the reported discrepancies and to arrive at a successful model of the air-shower development. The muon component of EAS provides not only a powerful key for primary mass measurement but also provides information regarding hadronic interactions.


\section*{Acknowledgments} 

The numerical simulations were carried out in laboratories created under the project
``Development of research base of specialized laboratories of public universities in Swietokrzyskie region'',
POIG 02.2.00-26-023/08, 19 May 2009.\\
MR was supported by the Polish National Science Centre (NCN) grant 2016/23/B/ST2/00692.

\appendix

\section{Strangelets in cosmic rays}\label{app:stran}

The analysis of the EAS data offers a unique possibility to observe possible imprints of strangelets arriving from the outer space. They are lumps of SQM, a new possible stable form of matter (cf.~\cite{Klingenberg:1999sb,Witten:1984rs,Alcock:1988re} for details). Following~\cite{Wilk:1996ba} it is fully reasonable to search 
for SQM in cosmic ray experiments because the specific features of strangelets~\cite{Alcock:1985vc,Farhi:1984qu} allow them to penetrate deep into the atmosphere~\cite{Wilk:1996je,Kankiewicz:2016dha}. The point is that there is a certain critical size of the strangelet, given by the critical value of its mass number $A = A_{crit} \simeq 300$~\textemdash~$400$, such that for $A > A_{crit}$ strangelets are absolutely stable against neutron emission. Below this limit strangelets decay rapidly evaporating neutrons. The spatial radii of strangelets turn out to be comparable to the radii of ordinary nuclei~\cite{Wilk:1996ba}, i.e., their geometrical cross sections are similar to the normal nuclear ones. 
To account for their strong penetrability one has to accept that strangelets penetrating deeply into the atmosphere are formed in many successive interactions with the air nuclei by the initially very heavy lumps of SQM entering the atmosphere and decreasing due to the collisions with air nuclei (until their $A$ reaches the critical value $A_{crit}$~\cite{Wilk:1996ba}). 
Such scenario is fully consistent with all present experiments~\cite{Wilk:1996ba,Wilk:1996je,Kankiewicz:2016dha,Wilk:1996me,GladyszDziadus:1997vw}. In this scenario the interaction of a strangelet of mass $A$ with an air nucleus of mass $A_{air}$ involves all quarks of the target located in the geometrical intersection of the colliding air nucleus and the strangelet. So, up to $3\cdot A_{air}$ quarks from the strangelet could be used making its mass drop to a value of $A-A_{air}$. The total penetration depth of the strangelet is then equal to
\begin{align}
\Lambda & \simeq \frac{1}{3}\lambda_{NA_{air}}\left( \frac{A}{A_{air}}\right)^{\frac{1}{3}}\left( 1 - \frac{A_{crit}}{A}\right)^2 \left( 4 - \frac{A_{crit}}{A} \right) \nonumber \\
& \simeq \frac{4}{3} \lambda_{NA_{air}}\left(\frac{A}{A_{air}}\right)^{\frac{1}{3}}, 
\label{eq:pen_depth}
\end{align}
where $\lambda_{NA_{air}}$ is the mean free path for $N-A_{air}$ interactions. This mechanism fully agrees with all present and proposed experiments and can be also checked in the future by measuring the products of the intermediate collisions.

There are several reports suggesting the existence of direct candidates for SQM~\cite{Saito:1990ju,Price:1976mr} (characterized mainly by their very small $Z/A$ ratios). All of them have mass numbers {\it A} near or slightly exceeding $A_{crit}$. Analysis of these candidates for SQM show~\cite{Wilk:1996ba,Wilk:1996je,Kankiewicz:2016dha} that the abundance of strangelets in the primary cosmic ray flux is $F_{SQM}(A = A_{crit})/F_{tot}\simeq 2.4\cdot 10^{-5}$ at the same energy per particle. For a normal flux of primary cosmic rays~\cite{Shibata:1995xz} the expected flux of strangelets is then equal to $F_{SQM}\simeq 7\cdot 10^{-6}~{\rm m}^{\rm -2} {\rm h}^{\rm -1} {\rm sr}^{\rm -1}$ for the energy above 10 GeV per initial strangelet.

Experimental results show a wide spectrum of exotic events (Centauros, superfamilies with 'halo', strongly penetrating component, etc.) which are clearly incompatible with the standard ideas 
of hadronic interactions known from the accelerator experiments. Some new mechanism or new primaries are therefore needed. Assuming that strangelets represent such new primaries 
one is able to explain~\cite{GladyszDziadus:1997vw} (at least to some extend) a strong penetrating nature of some {\it abnormal} cascades associated with their very slow attenuation and with the appearance of many maxima with small distances between them (about 2 - 3 times smaller than in the {\it normal} hadron cascades). The already mentioned Centauro (and mini-Centauro) events, characterized by the extreme imbalance between hadronic and gamma-ray components among produced secondaries, are probably the best known examples of such exotic events. They require a deeply penetrating component of the cosmic rays. We claim that they can be a product of strangelets penetrating deeply into atmosphere and evaporating neutrons~\cite{Wilk:1996je}. Both the flux ratio of Centauros registered at different depths and the energy distribution of secondary particles within them can be
successfully described by such concept.

Anomalous events have been reconfirmed by measuring extensive air showers cf., for example,~\cite{Shaulov:1996zy}. Among them was the striking observation~\cite{Antonova:1999xw} of extremely long-delayed neutrons in connection with the large EAS which cannot be explained by the standard mechanism of hadronic cascade development. Also muon bundles of extremely high multiplicity, observed recently by ALICE~\cite{ALICE:2015wfa} and ALEPH~\cite{Avati:2000mn} detectors in their dedicated cosmic-ray run, can originate from collisions of strangelets with the atmosphere~\cite{Rybczynski:2000vc,Kankiewicz:2016dha}.

The experimental data mentioned above lead to a flux of strangelets which is consistent with the astrophysical limits and with the upper limits given by the experiment~\cite{Sahnoun:2008mr}. The data follows the $A^{-7.5}$ behaviour, which coincides with the behaviour of abundance of normal nuclei in the Universe~\cite{Rybczynski:2001bw}. Interpretation of indirect observations (anomalous events observed in emulsion chambers, and also results from the measurements of EAS) can provide signals of strangelets. Moreover, the direct identification (by implementing passive nuclear track detector arrays) of SQM is quite realistic in the nearby future. All these considerations motivate further experimental search for the SQM and for its cosmological and elementary particle physics aspects.

\section{Model description}\label{app:model_desc}

In our simulations we used the F00 model which consists in a simple non-scaling extrapolation of the inclusive data at ISR and SPS energies~\cite{Wrotniak1985}. In this version, the leading particle remembers its charge, and the $x$-distribution of secondaries does not depend on the elasticity (relative energy carried by the leading particle). For collisions with air nuclei, the inelastic cross-section dependence on energy is given by:
\begin{eqnarray}
\sigma_{inel} = \sigma_{0} + \delta \cdot \left[\ln\left(E/E_{0}\right)\right]^{\alpha},~~~{\rm for}~E\geq E_{0}\\
\sigma_{inel} = \sigma_{0},~~~{\rm for}~E<E_{0}.
\label{eq:sigma}
\end{eqnarray}
Numerical values of the parameters are quoted in table~\ref{tab:param_sigma}.
\begin{table}
\tbl{Parameters of the energy dependence of inelastic cross-section given by equation~\ref{eq:sigma} used in SHOWERSIM simulations.}
{\begin{tabular}{@{}ccccc}
\hline\noalign{\smallskip}
Particle & $\sigma_{0}$~[mb] & $\delta$~[mb] & $E_{0}$~[TeV] & $\alpha$\\
\noalign{\smallskip}\hline\noalign{\smallskip}
proton,neutron & 280 & 2.5 & 0.1 & 1.8\\
pion & 196 & 1.7 & 0.067 & 1.8\\
kaon & 178 & 1.6 & 0.067 & 1.8\\
\noalign{\smallskip}\hline
\end{tabular}\label{tab:param_sigma}}
\end{table}

\begin{figure*}[t]
\begin{center}
\includegraphics[angle=0,width=0.8\textwidth]{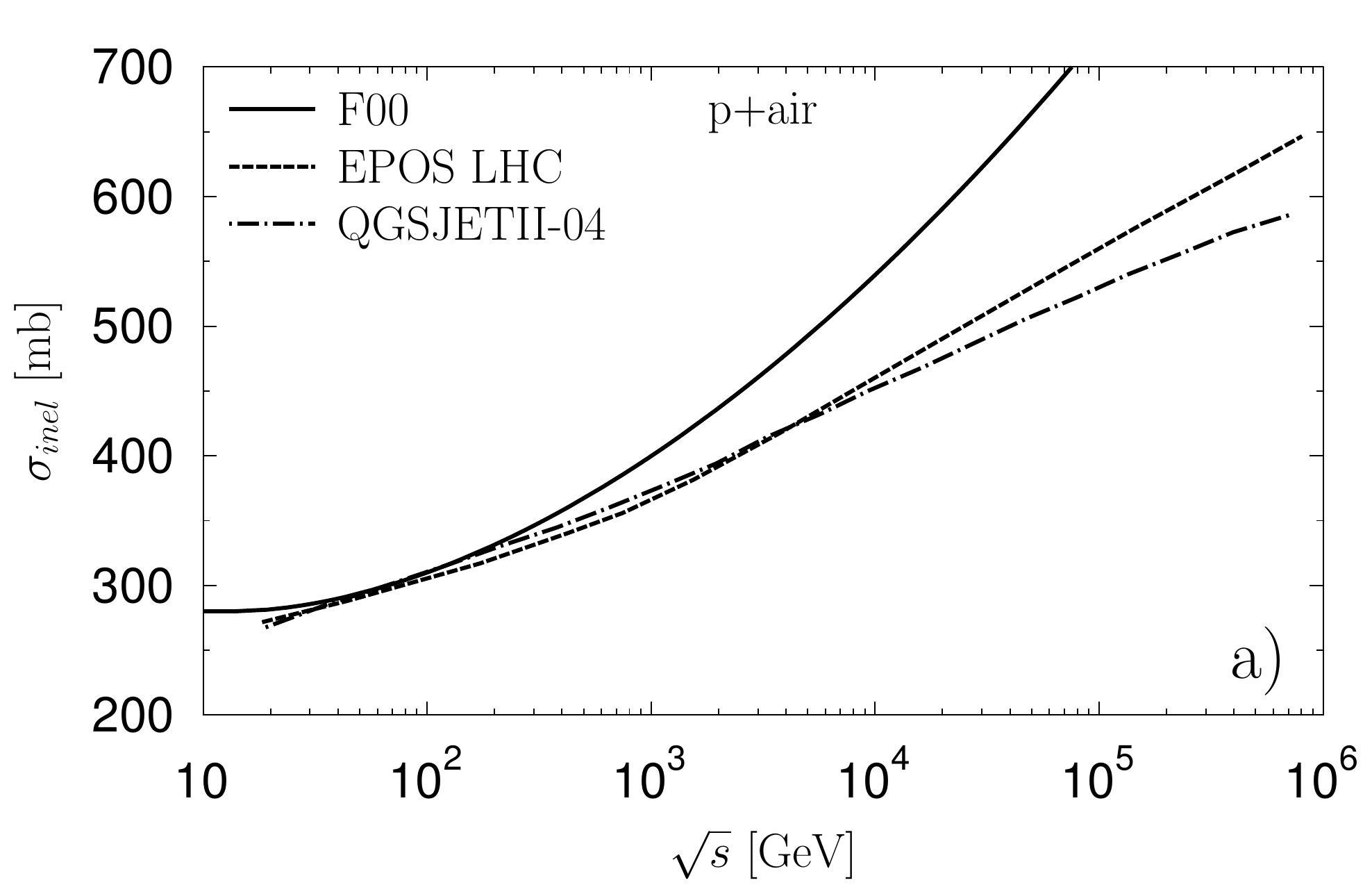}\\
\includegraphics[angle=0,width=0.8\textwidth]{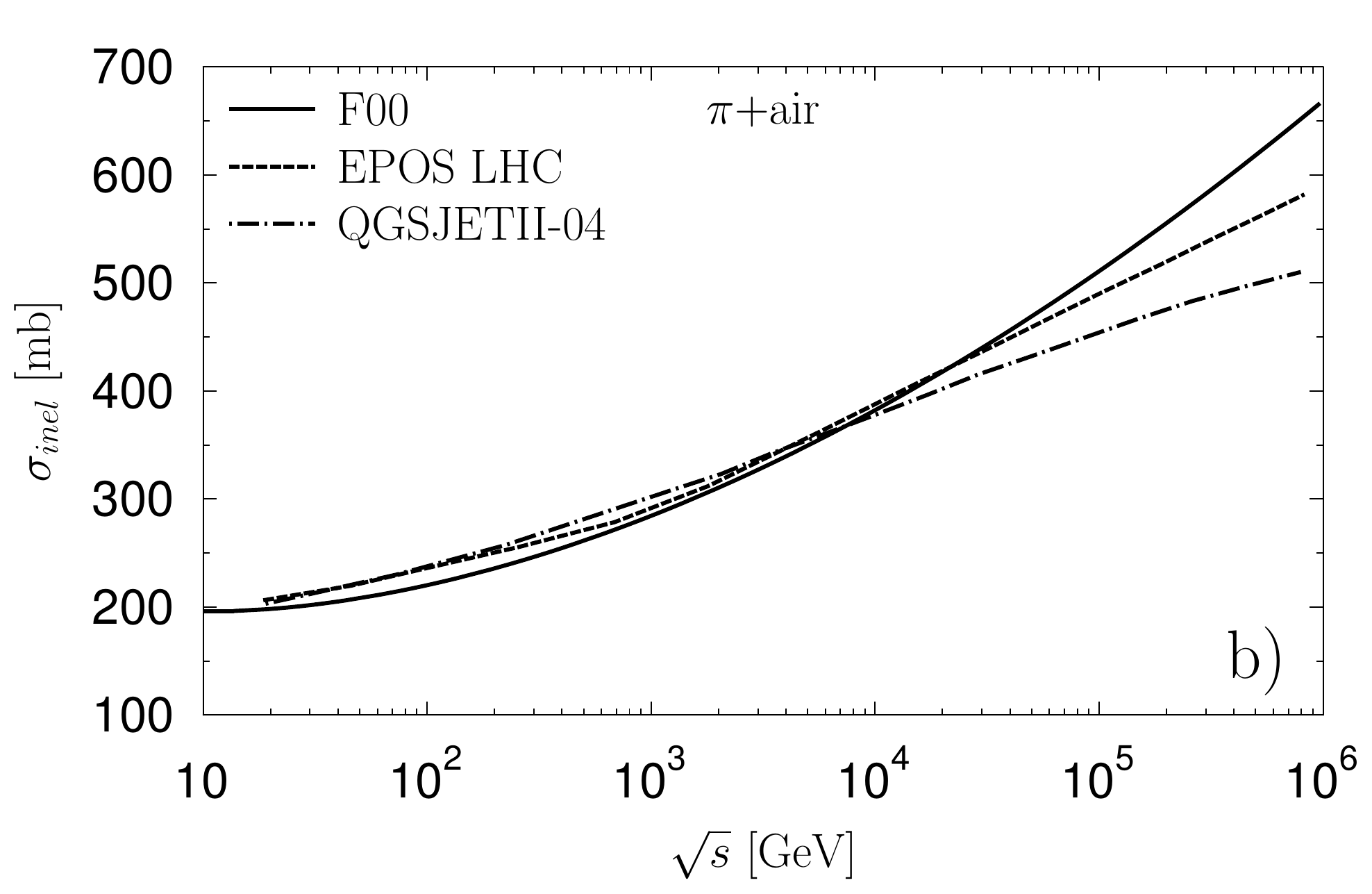}
\end{center}
\caption{\small Energy dependence of the inelastic cross-section in proton-air (panel a) and pion-air (panel b) interactions for our F00 model (full line). EPOS LHC (dotted line) and QGSJETII-04 (dash-dotted line) predictions are taken from~\cite{Pierog:2017nes}.}
\label{fig:sigma_inel}
\end{figure*}
 
In figure~\ref{fig:sigma_inel} we show the energy dependence of the inelastic cross-section in proton-air (panel a) and pion-air (panel b) interactions for our F00 model. For comparison we also show the EPOS LHC and QGSJETII-04 predictions taken from~\cite{Pierog:2017nes}. Note that the resulting ratios of cross sections in our F00 model do not vary significantly with the energy.

\begin{figure}[t]
\begin{center}
\includegraphics[angle=0,width=0.8\textwidth]{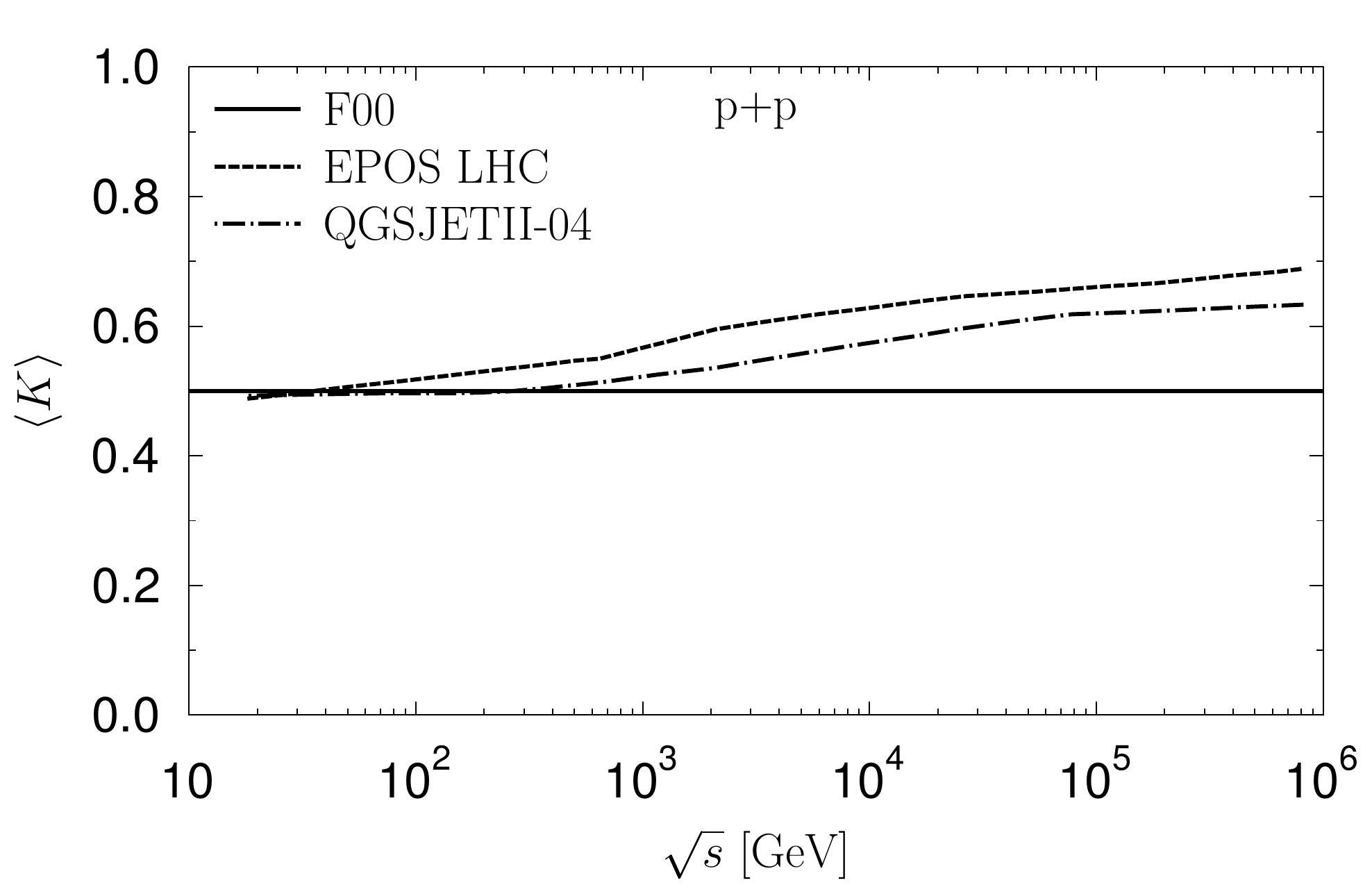}
\end{center}
\caption{\small Average inelasticity in proton-proton interactions as a function of the center-of-mass energy for our F00 model (full line). EPOS LHC (dotted line) and QGSJETII-04 (dash-dotted line) predictions are taken from~\cite{Pierog:2017nes}.}
\label{fig:inelast}
\end{figure}

In the F00 model the inelasticity $K$, i.e. the relative energy used for production of secondaries, is sampled from a uniform distribution over an (0,1) interval for interacting nucleons and over an (0.333,1) interval for meson interactions. Therefore, the average inelasticity $\langle K\rangle=0.5$ for nucleons and $\langle K\rangle=0.667$ for mesons. In figure~\ref{fig:inelast} we show the average inelasticity in proton-proton interactions as a function of the center-of-mass energy for our F00 model (full line). EPOS LHC (dotted line) and QGSJETII-04 (dash-dotted line) predictions are taken from~\cite{Pierog:2017nes}.

\begin{figure}[t]
\begin{center}
\includegraphics[angle=0,width=0.8\textwidth]{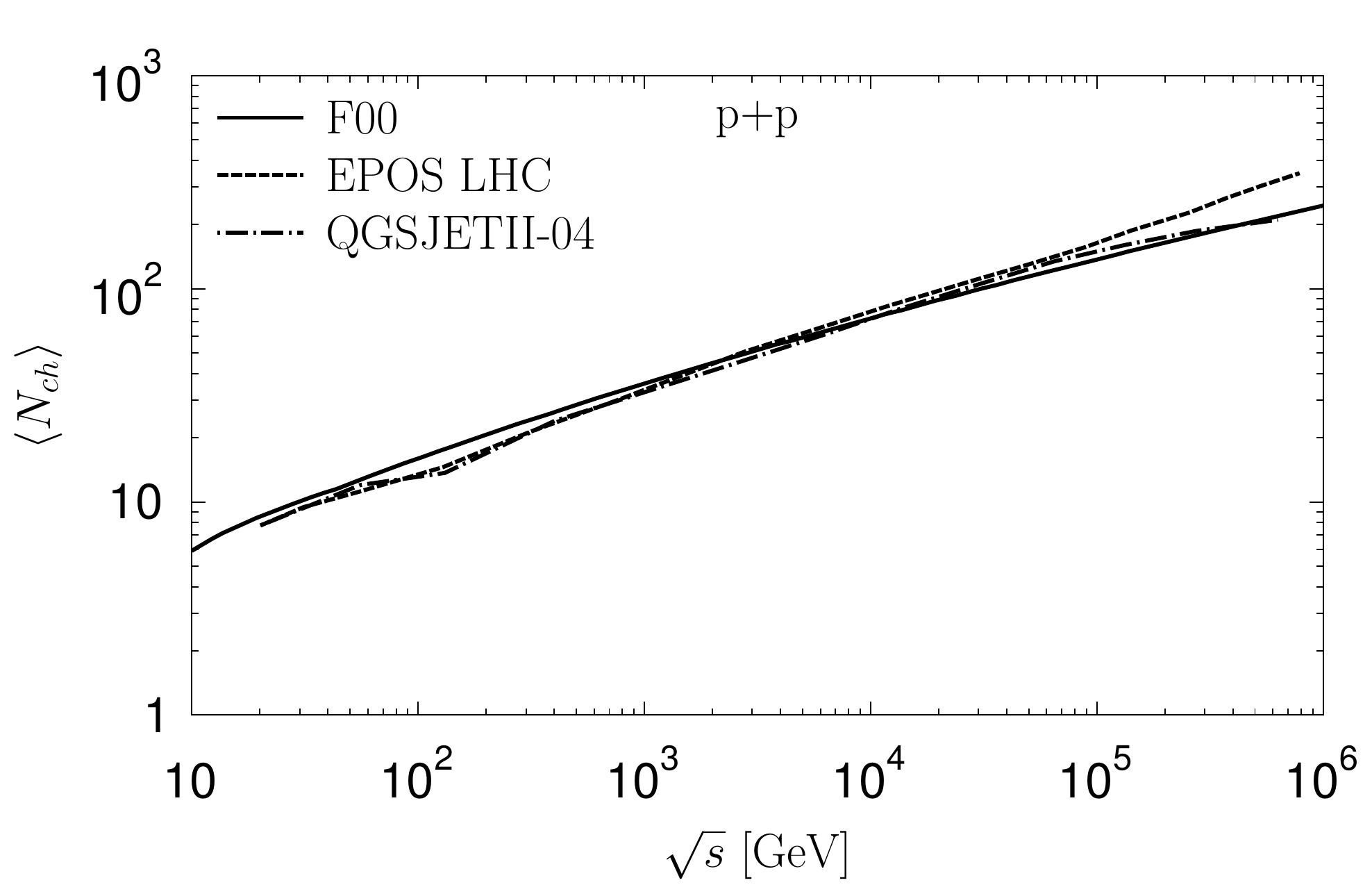}
\end{center}
\caption{\small Mean multiplicity of charged particles produced in proton-proton interactions as a function of center-of-mass energy for our F00 model (full line). EPOS LHC (dotted line) and QGSJETII-04 (dash-dotted line) predictions are taken from~\cite{Pierog:2017nes}.}
\label{fig:mult_kpi}
\end{figure}

\begin{figure}[t]
\begin{center}
\includegraphics[angle=0,width=0.8\textwidth]{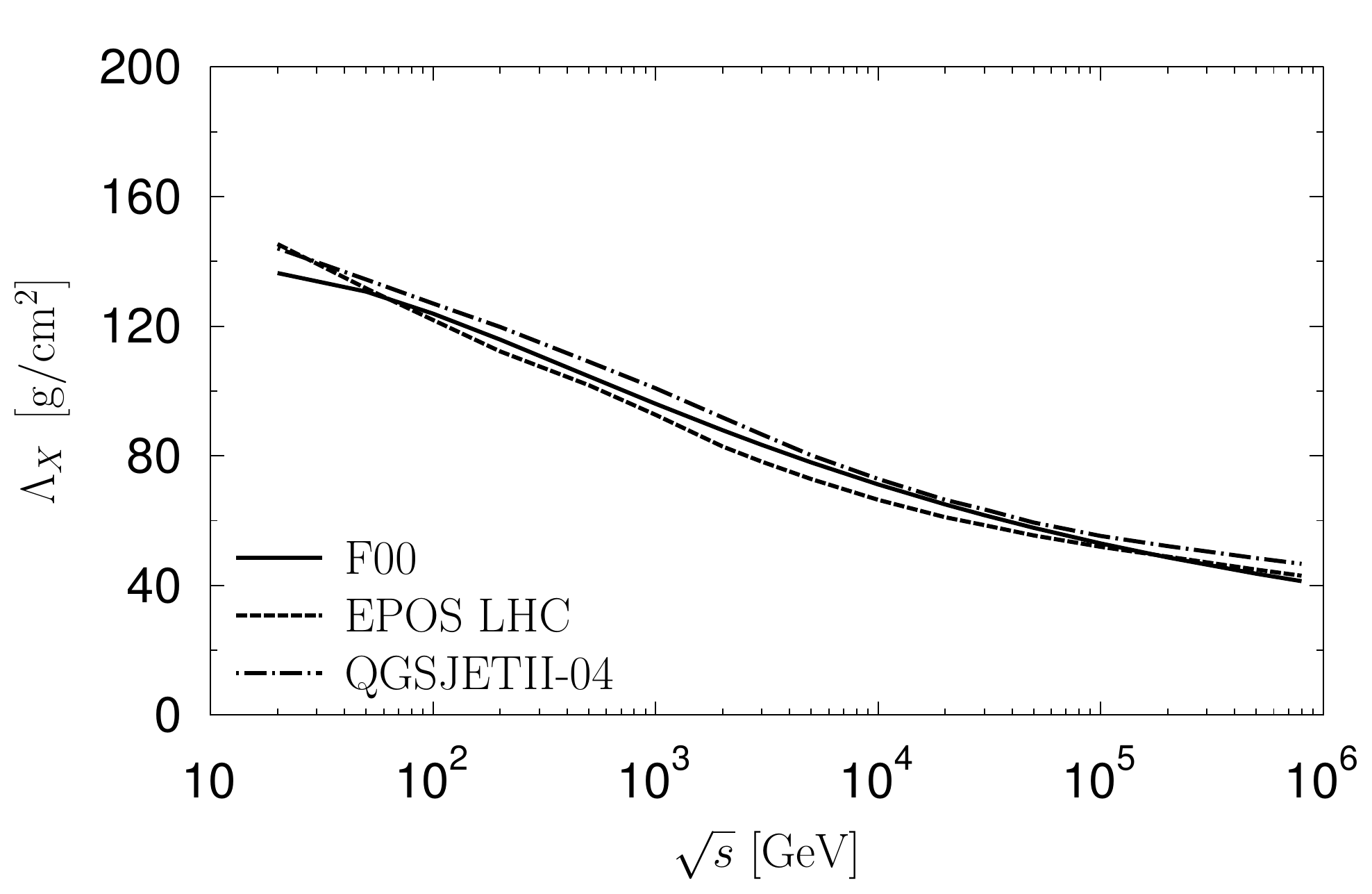}
\end{center}
\caption{\small Attenuation length as a function of the center-of-mass energy for our F00 model (full line). EPOS LHC (dotted line) and QGSJETII-04 (dash-dotted line) predictions are taken from~\cite{Pierog:2017nes}.}
\label{fig:atten}
\end{figure}

In figure~\ref{fig:mult_kpi} we show the mean multiplicity of charged particles produced in proton-proton interactions as a function of the center-of-mass energy for our F00 model (full line). Again, EPOS LHC (dotted line) and QGSJETII-04 (dash-dotted line) predictions are taken from~\cite{Pierog:2017nes}. 

In the F00 model the identity of secondary particle is generated with the probability
\begin{equation}
p\left(x,E\right) = \alpha\left(x\right)  + \beta\left(x\right)\ln\left(E\right),
\label{eq:F00_ident_gen}
\end{equation}
where $x=p,\pi,K$ denotes various secondaries at 1~TeV. Probability of the emission of charged particle ($\pi^{\pm}, K^{\pm}$) is:
\begin{align}
p\left(N_{ch},E\right) & = 1-0.5\left[\alpha\left(K\right)+\beta\left(K\right)\ln\left(E\right)\right]\nonumber \\ 
& - \alpha\left(\pi^{0}\right)-\beta\left(\pi^{0}\right)\ln\left(E\right).
\label{eq:F00_ident_part}
\end{align}
In the standard version of F00 model $\alpha\left(\pi^{0}\right)=0.3$, $\alpha\left(K\right)=0.1$, $\beta\left(\pi^{0}\right)=0.0128$, and $\beta\left(K\right)=0.0028$. The fraction of produced neutral pions increases by 0.0295 per energy decade and fraction of produced kaons increases by 0.006 per decade. Changing the value of parameter $\beta$ we can obtain a different number of charged particles, $\langle N_{ch}\rangle$ for a fixed number of secondaries, $\langle N\rangle=\rm{const}$.

In the development of EAS, the inelasticity $K$ and the cross section for interactions $\sigma$ are strongly correlated. The attenuation of hadrons or the depth of the shower maximum $X_{max}$ are actually the measure of combinations of $K$ and $\sigma$, and the effect of these two parameters is extremely difficult to disentangle. From the above we can see that $\langle K\rangle$ and $\sigma_{inel}$ differ significantly for different models. Nevertheless, the $\langle X_{max}\rangle$ for these models are quite similar. 

The deep tail of the depth of maximum distribution, which has an exponential behaviour:
\begin{equation}
\frac{dN}{dX_{max}} \sim \exp \left(-\frac{X_{max}}{\Lambda_X} \right) 
\label{eq:Xdistr}
\end{equation}
depends on the proton interaction length $\lambda = 14.45~m_{p} / \sigma_{inel} = 2.4\cdot 10^4 /\sigma_{inel}$ $\left[{\rm g/cm^2}\right]$ via shower maxima attenuation length:
\begin{equation}
\Lambda_X \simeq 0.8\frac{2.4\cdot 10^4}{\langle K\rangle \sigma_{inel}}~~~\left[{\rm g/cm^2}\right].
\label{eq:LamX}
\end{equation}
Figure~\ref{fig:atten} shows the shower maxima attenuation length as a function of the center-of-mass energy for our F00 model (full line). EPOS LHC (dotted line) and QGSJETII-04 (dash-dotted line) predictions are taken from reference~\cite{Pierog:2017nes}. Despite the very different energy dependence of $\langle K\rangle$ and $\sigma_{inel}$, the shower maxima attenuation lengths $\Lambda_X$ are very similar.

We evaluated the attenuation of the showers maxima, $\Lambda_{X}$, for protons at $10^{18}$~eV  obtaining $\Lambda_{X}=58.5\pm 1~{\rm g/cm^{2}}$, which agrees nicely with the experimental data $\Lambda_{X}=55.8\pm 2.3~{\rm g/cm^{2}}$ at $10^{18.2}$~eV, reported by the Pierre Auger Collaboration in~\cite{Collaboration:2012wt}, $\Lambda_{X}=57.4\pm 1.8~{\rm g/cm^{2}}$ at $10^{18}$~\textemdash~$10^{18.5}$~eV, and $\Lambda_{X}=60.7\pm 2.1~{\rm g/cm^{2}}$ at $10^{17.8}$~\textemdash~$10^{18}$~eV, given by the Telescope Array experiment~\cite{Abbasi:2016xxr}.


\end{document}